\documentclass[12pt]{article}
\input{psfig}
\usepackage{array}
\usepackage{delarray}
\newcommand{\be}{\begin{equation}}
\newcommand{\ee}{\end{equation}}
\newcommand{\ba}{\begin{array}}
\newcommand{\ea}{\end{array}}
\title{{\bf D-BRANES, MONOPOLES AND NAHM EQUATIONS}}
\author{Duiliu-Emanuel Diaconescu\\
        {\it Department of Physics and Astronomy }\\
           {\it Rutgers University}\\
           {\it Piscataway, NJ 08855-0849}\\
           {\small e-mail: duiliu@physics.rutgers.edu}}
\begin{document}
\maketitle
\thispagestyle{empty}
\begin{abstract}
We study the correspondence between IIb solitonic 1-branes and 
monopoles in the context of the 3-brane realization 
of $D=4$ $N=4$ super Yang-Mills theory. We show that a bound state of
1-branes stretching between two separated 3-branes exhibits a family 
of super-symmetric ground states that can be identified with the ADHMN
construction of the moduli space of $SU(2)$ monopoles.. This identification is
supported by the construction of the monopole gauge field as a
space-time coupling in the 
quantum mechanical effective action of a 1-brane used as a probe. 
The analysis also reveals an intriguing aspect of the 1-brane theory:
the transverse oscillations of the
1-branes in the ground states are described by non-commuting matrix
valued fields which develop poles at the boundary.
Finally, the construction is generalized to $SU(n)$ monopoles with
arbitrary $n>2$.
\end{abstract}
\clearpage

\section{Introduction}

It has been argued recently [11],[12],[31],[33],[35] that the
$SL(2,\,Z)$ duality of the $D=4$, $N=4$ super-symmetric Yang-Mills
theory can be viewed as the field-theoretic
counterpart of the more fundamental $SL(2,\,Z)$ duality of type IIb
super-string theory [30]. A precise formulation of this correspondence 
can be achieved in the context of Dirichlet branes of super-string
theory, whose existence and basic properties have been first shown
in [7],[21],[27],[26]. Especially, $D=4$, $N=4$ super-symmetric Yang-Mills
theory with gauge group $SU(2)$ is realized as the low energy
effective theory of two parallel type IIb 3-branes. A configuration
with separated 3-branes corresponds to a point of spontaneously symmetry
breaking on the moduli space of the theory, the scale of the Higgs
mechanism being essentially proportional to the distance between the branes.
The excitations of this system consist of charged W-bosons which can
be identified with fundamental open strings stretching between the
3-branes and of magnetic monopoles which can be similarly identified with 
solitonic strings also stretching between the branes. More generally,
the dyonic states of the world-volume Yang-Mills theory correspond to 
bound states of solitonic and fundamental strings. 
This correspondence is compatible with $SL(2,\,Z)$ transformations.

The present paper gives an explicit construction of the moduli space
of world-volume monopoles as a moduli space of super-symmetric ground
states of solitonic strings. This construction relies on the
remarkable description of solitonic excitations of super-string theory
as Dirichlet branes [27].
The relevant formulation of the monopole moduli spaces turns out 
to be the that given by Nahm in [24], usually called ADHMN
construction. In this respect our results are the monopole counterpart
of the similar constructions carried out in [9],[10],[36],[37] in the
framework of instanton moduli spaces.

We first give a detailed analysis for $SU(2)$ gauge group 
emphasizing the role of boundary conditions. A probe analysis along
the lines of [9] proves to be crucial for a proper understanding of
the correspondence. In the last section we generalize the construction
to $SU(n)$ gauge groups with $n>2$.

{\bf Acknowledgments}
I am deeply grateful to Michael R. Douglas for introducing me to these
problems as well as for constant encouragement and support. Without
his patient supervising and enlightening suggestions this work would
have never been possible. I thank Roger Bielawski for an enlightening 
e-mail discussion and for making his work [4] available to me prior to
publication.

\section{D-Branes and SU(2) Monopoles}
\subsection{\it{SU(2) Monopoles and Nahm Equations}}

The data for an $SU(2)$ monopole on $R^3$ consist of a connection
$A_{\mu}(x)$ in the trivial bundle $SU(n)\times{R^3}$ and a Higgs
field $\Phi(x)$ transforming in the adjoint representation of
$SU(n)$. The energy of the Yang-Mills-Higgs functional has a lower topological
bound which is attained by certain static field configurations, the
BPS monopoles. In the limit of vanishing Higgs potential these are
defined by Bogomolnii equations:
\be
B_i=D_i\Phi\mbox{,}\qquad    B_i={1\over{2}}\epsilon_{ijk}F_{jk}
\ee
with asymptotic boundary conditions on $\Phi$:         
\be
\Phi(r)=i\mbox{diag}(1,\ -1)-{i\over{2r}}\mbox{diag}
(k,\ -k)+O({1\over{r^2}}) 
\ee
The expectation value of $\Phi$ breaks the gauge group $SU(2)$ down to
an electro-magnetic $U(1)$ in the asymptotic region. Moreover, it
obviously determines a map from the two sphere at infinity
$S^2_\infty$ into the orbit $SU(2)/U(1)\cong S^2$ which defines in fact a
homotopy class in $\pi_2(S^2)$. This class yields a topological
invariant of the solution. In this case $\pi_2(S^2)\cong Z$ and the
topological invariant is given simply by $k\in Z$, also called
magnetic charge.

A remarkable description of the moduli space of monopoles
for arbitrary (classical) gauge group has been given by Nahm in [24] 
and further developed in [8], [15], [18]. We will quote their results only for
the moduli space ${\cal M}(k)$ of $SU(2)$ monopoles with fixed
magnetic charge $k> 0$. A complex of Nahm data consists of $su(2)$
valued functions $X^i$, $i=1,2,3$ on the interval $(-1,\ 1)$
satisfying Nahm equations:
\be
{d{{X^i}}\over{ds}}+{1\over 2}{\sum_{j{,}k=1}^3}\epsilon_{ijk}[{X^j},
{X^k}]=0  
\ee
subject to the following boundary conditions:

$(i)$ The $X^i$ are analytic in the interior of the interval with simple
poles at $s=\pm 1$:
\be
X^i={T^i\over{s\mp 1}}+O(1),\qquad s\mapsto\pm 1
\ee
$(ii)$ The residues $T^i$ define an irreducible $k$ dimensional
representation of $SU(2)$:
\be
[T^i,\ T^j]=\epsilon^{ijk}T^k.
\ee
Then there is a $1-1$ correspondence between $U(k)$ conjugacy classes of Nahm
complexes and $SU(2)$ monopoles of charge $k$. This construction can
be further refined [15] by imposing a symmetry condition on the
functions $X^i$:
\be
{X^i}(s)={X^i}^\top(-s)
\ee
and restricting to $O(k)$ conjugation.

Nahm equations can be set in covariant form [8] by introducing a
fourth component $X^4$ and defining the covariant derivative:
\be
\nabla_s{X^i}={{d{X^i}\over{ds}}+[{X^0}{,}{X^i}]}
\ee
Then they prove to be equivalent to self-duality equations for the
connection ${X^0}{ds}+{X^i}{d{x^i}}$ on the space $(s,{x^i})$:
\be
\nabla_s{X^i}+{1\over 2}\sum_{j{,}k=1}^3{\epsilon_{ijk}}[{X^j}{,}{X^k}]=0
\ee
The covariant Nahm data are invariant under unitary gauge
transformation which restrict to the identity transformation at 
$s=\pm 1$:
\be
X^i\mapsto{g^{-1}{X^i}g}\qquad    
X^0\mapsto{g^{-1}{X^0}g+g^{-1}{dg\over{ds}}}
\ee
According to [8] one can further define complex Nahm data:
\be
\alpha={1\over{2}}({X^0}+i{X^1}){,}\qquad
\beta={1\over{2}}({X^2}+i{X^3})
\ee
in terms of which the original equations become a complex equation:
\be
{d{\beta}\over{ds}}+2[\alpha{,}\beta]=0
\ee
and a real equation:
\be
{d\over{ds}}(\alpha+{\alpha}^{*})+2([\alpha{,}{\alpha}^{*}]+[\beta{,}{\beta}^{*}])=0
\ee
The complex equation is preserved by complex $GL(m,C)$-valued
gauge transformations:
\be
\alpha\mapsto{g^{-1}}{\alpha+{1\over{2}}{g^{-1}}{d\over{ds}}g}\qquad
  \beta\mapsto{{g^{-1}}{\beta}g}
\ee
while the real equation is preserved only by unitary
transformations.
The equations (11) and (12) present a striking resemblance with the
moment map equations for hyperkahler quotients [10]. Actually, it can
be rigorously proven [19] that the moduli space arises in this
formulation as a hyperkahler quotient.  

Finally we briefly review the construction of the monopole solution in
terms of Nahm data [2], [6], [15], [24]. Consider the first order
linear differential operator:
\be
\Delta (s)=i\frac{d}{ds}+(T^i+ix^i)\sigma_i
\ee
where $\sigma_i$ are the standard generators of $SU(2)$. Then the
equation
\be
\Delta^\dagger v=0
\ee
has a unique normalizable solution $v(s)$ with
\be
\int\, ds\, v^\dagger v=1
\ee
Note that this is a quaternionic $1\times k$ matrix which in complex
notation becomes a complex $2\times 2k$ matrix, [4], [11].The gauge
field and the Higgs field of the monopole are given
respectively by:
$$
A_i=\int\, ds\, v^\dagger\partial_i v
$$
\be
\Phi=-i\int\, ds\, sv^\dagger v
\ee
This concludes our brief discussion of $SU(2)$ monopoles.

\subsection{\it{Moduli Space of Super-Symmetric Ground States}}

Consider two parallel Dirichlet type IIb 3-branes in the $123$ plane
separated by a distance $\mu$ along the $x^9$ axis and a bound state of $k$
1-branes stretching between them. 
The existence of such bound states
has been proven in [35], while the fact that they can end on 3-branes
has been shown in [31],[12],[32]. As stated in [7], [21], [27] and
emphasized further in [26], [35] the low energy effective action
describing the 1-brane dynamics is the dimensional reduction of the 
$D=10$, $N=1$ super Yang-Mills action to the 1-brane world-sheet. 
The resulting action has $(8,8)$ super-symmetry and it has been derived
in [5]. The field content and it's D-brane interpretation can be
summarized as follows:

a)bosonic sector
$$A_{\mu}\qquad{\mu=0,1}$$
$${\Phi_{A4}}={(X_{A}+iX_{A+3})}/\sqrt{2}\qquad {A=1,2,3}$$
$${\Phi^{AB}}={1\over{2}}{\epsilon}^{ABCD}{\Phi_{CD}}={{\Phi}_{AB}}^*\qquad{A,B=1{,}\ldots{,}4}$$
$$S=X_7{,}\qquad  P=X_8$$

b)fermionic sector
$$\chi^A{,}\qquad  {\tilde{\chi}^A}=C_2{{({\bar{\chi}^A})}^T}\qquad
{A=1{,}\ldots{,}4}$$
In the above $A_{\mu}$ is the two dimensional gauge field,
$\Phi_{AB}$, $S$, $P$ are two dimensional scalars in the adjoint
representation of $U(m)$ that represent the transverse oscillations of
the $1$-branes and $\chi^A$ are fermions transforming also in the
adjoint of $U(m)$. The indices $A$, $B$, $\ldots$ are $SU(4)$ symmetry
indices arising in the process of dimensional reduction and the fields
$\Phi_{AB}$ form an antisymmetric tensor multiplet of $SU(4)$. We follow the
conventions of [5] for the two dimensional Dirac algebra. The
lagrangian and super-symmetry transformations read 
$$
{\cal L}
=Tr\{-{1\over{4}}F_{\mu\nu}F^{\mu\nu}+{1\over{2}}D_{\mu}PD^{\mu}P+{1\over{2}}D_{\mu}SD^{\mu}S
$$
$$
+{1\over{2}}D_{\mu}{\Phi_{AB}}D^{\mu}{\Phi^{AB}}
+i{\bar{\chi}_A}{\gamma}\cdot{D}{\chi}^A+g({\bar{\chi}_A}{\gamma}_5[{\chi}^A{,}P]+i{\bar{\chi}_A}[{\chi}^A{,}S])
$$
$$
-{1\over{2}}ig{({\bar{\tilde{\chi}}}^A[{\chi}^B{,}{\Phi}_{AB}]\\-{\bar{\chi}}_A[{\tilde{\chi}}_B{,}{\Phi}^{AB}])}
-{1\over{4}}{g^2}[{\Phi}_{AB},{\Phi}_{CD}][{\Phi}^{AB},{\Phi}^{CD}]-{1\over{2}}{g^2}[S,P]^2
$$
\be
-{1\over{2}}{g^2}[S,{\Phi}_{AB}][S,{\Phi}^{AB}]-{1\over{2}}{g^2}[P,{\Phi}_{AB}][P,{\Phi}^{AB}]\}
\ee
$$
\delta A_\mu=i{(\bar{\alpha}_A\gamma_\mu\chi^A-\bar{\chi}_A\gamma_\mu\alpha^A)}
$$
$$
\delta P=\bar{\chi}_A\gamma_5\alpha^A-\bar{\alpha}_A\gamma_5\chi^A
$$
$$
\delta S=i{(\bar\chi_A\alpha^A-\bar\alpha_A\chi^A)}
$$
$$
\delta\Phi_{AB}=i{(\bar\alpha_B\tilde\chi_A-\bar\alpha_A\tilde\chi_B+\epsilon_{ABCD}{\bar{\tilde{\alpha}}}^C\chi^D)}
$$
$$
\delta\chi^A=\sigma_{\mu\nu}F^{\mu\nu}+i\gamma\cdot
DP\gamma_5\alpha^A-\gamma\cdot DS\alpha^A -\gamma\cdot D\Phi^{AB}\tilde\alpha_B
$$
\be
+g(i[P,S]\gamma_5\alpha^A-i[P,\Phi^{AB}]\gamma_5\tilde\alpha_B-[S,\Phi^{AB}]\tilde\alpha_B+{1\over
2}[\Phi^{AB},\Phi_{BC}]\alpha^C)
\ee

Throughout this section we normalize the Higgs field so that $\mu=2$,
thus the 3-branes can be taken at points $\pm 1$ on the $x^9$ axis
without loss of generality. Let $s\times{t}\in(-1,\ 1)\times R$ be
world-sheet coordinates. Since the 1-branes are constrained to end
on the 3-branes one should obviously impose the boundary conditions
\be
{X^{\mu}}(\pm{1},t)=0{,}\qquad \mu=4{,}\ldots{,}8
\ee
Compatibility with super-symmetry transformations implies similar
boundary conditions for fermions:
\be
{\chi^A}(\pm{1},t)=0{,}\qquad A=1{,}\ldots{,}4
\ee
Since the super-symmetry transformations of the spinor fields involve
derivatives of the fields $X^{\mu}$ a simple consistency check of the
vanishing order near the boundary shows that we have in fact to
restrict to fields with compact support inside the interval or ''bump
fields''. This restriction does not apply to the transverse fields
$X^i$, with $i=1{,}2{,}3$ which will be seen to have interesting
boundary behavior. Moreover since we are interested in super-symmetric
ground states with $X^{\mu}$ and ${\chi^A}$ vanishing identically this
restriction is quite natural. There is a slight subtlety related to
this picture: since the world-sheet has nonempty boundary, the total
derivative terms in the super-symmetric variation of the lagrangian
yield surface terms by integration. However these terms cancel by the
above boundary conditions leading to a consistent theory. There are
also additional fields arising from the quantization of the
fundamental $1-3$ and $3-1$ strings. These constitute quantum
mechanical degrees of freedom that couple to the world-sheet boundary
and they will play an important role latter.

We can now address the problem of super-symmetric ground states for the 
1-brane configuration. These are solutions to:
\be
\delta \chi^A=0,\qquad{A=1,\ldots,4}
\ee
with:
\be
\chi^A\equiv 0,\qquad X^\mu\equiv 0,\qquad\mu=4,\ldots,8
\ee
Fixing the axial gauge $A_0=0$, which restricts us to static gauge 
transformations and imposing a reality condition on the super-symmetry
parameters:
\be
 \tilde\alpha^A_\varrho=\epsilon_{\varrho\gamma}\alpha^A_\gamma
\ee
we find a family of ground states that break half of the original
$(8,\ 8)$ supersymmetries given by:
\be
D_1\Phi^{AB}+{g\over 2}[\Phi^{AC},\ \Phi_{CB}]=0.
\ee
Setting $g=\sqrt 2$, the equations can be rewritten as:
\be
D_1X^i+{1\over 2}\epsilon^{ijk}[X^j,\ X^k]=0
\ee
which are formally identical to covariant Nahm equations (8). This is 
positive evidence for the identification of monopoles with 1-strings 
but it is by no means sufficient. There are two main problems that
have to be answered at this stage.

$(i)$ So far we have proceeded formally, ignoring the boundary
conditions on the fields $X^i(s)$ which are in
fact an essential ingredient of Nahm construction. Consequently, it is
vital that we understand these boundary conditions in D-brane context.

$(ii)$The second puzzle is related to the role of the usual
super-symmetric ground states. It can be seen easily that if the
reality conditions (24) are absent, the 1-brane system has a ``trivial'' family
of ground states:
\be
X^i=\mbox{diag}(i{\lambda^i}_1,\ldots,i{\lambda^i}_m)
\ee
which admit a physical interpretation in terms of positions of the
1-branes [35]. Thus one would be tempted to conclude
that these are the ``real'' ground states of the system while those
derived above are rather unphysical. As we show shortly these two problems
are in fact closely related. 

Note that the boundary conditions (20), (21) require that
$\delta\chi^A=0$ be identically satisfied in a neighborhood of the
boundary, thus the fields $X^i$ should behave near boundary as general
local solutions of Nahm equations. One can easily check that the
ansatz below:
\be
X^i={{T^i}\over s},\qquad [T^i,\ T^j]=i\epsilon^{ijk}T^k
\ee
always constitutes such a solution, thus the most general boundary conditions
allowed by consistency requirement are:
\be
X^i={{T^i}\over {s\mp 1}} +O(1), s\mapsto \pm{1}
\ee
with
\be
[T^i,\ T^j]=i\epsilon^{ijk}T^k
\ee
defining an $k$ dimensional $SU(2)$ representation. These are not yet
Nahm boundary conditions as the latter also require that the $SU(2)$
representation be irreducible. It is this aspect that will provide a
hint on the solution to the second problem as well. 

Suppose that the 
representation is reducible, decomposing as:
\be
k=k_1\oplus k_2\oplus\ldots\oplus k_q
\ee
Then the residues $T^i$  may be set in block diagonal form:
\be
T^i=\mbox{diag}({T^i}_1,\ldots{T^i}_q) 
\ee
Using a similar block diagonal ansatz for the fields $X^i(s)$:
\be
X^i=\mbox{diag}({X^i}_1,\ldots{X^i}_q)
\ee
we find that the Nahm equations split in $q$ groups 
\be
\frac{d{X^i}_a}{ds}+{1\over 2}\epsilon^{ijk}[{X^j}_a,\
{X^k}_a]=0\qquad a=1,\ldots,q
\ee
In D-brane terms this means that the original bound state of $k$
branes splits in $q$ sub-bound states infinitely far apart. Each of
the above equations determines a ground state for each group taken separately.
The extreme limit is the case when the representation
defined by the residues splits in a sum of one dimensional
representations. In this case the residues vanish and the solutions to
Nahm equations following from the ansatz are exactly the ``trivial''
ground states (27).

Collecting all the facts, we have shown that the bound state of $k$
1-branes exhibits a family of ground states which may be formally
identified with the monopole moduli space in Nahm formulation. This 
identification is achieved if one imposes Nahm boundary conditions on 
the transverse fields $X^i(s)$. The usual flat directions of the
world-sheet potential appear as degenerate Nahm solutions
corresponding to the limit of infinitely separated 1-branes. However,
it is not clear why Nahm boundary conditions are the right ones
when the D-branes are close by. Their physical interpretation is
obscured by the poles of the transverse fields at the boundary and by
the fact that in general the matrix valued ``coordinate'' fields 
cannot be simultaneously diagonalized as
they do not commute. One could argue that the usual ground states
should be the ``real'' ones for any D-brane configuration. It appears that
these questions can be clarified by a probe argument.

\subsection{\it Probe Analysis}

The main idea is to use a D-brane as a probe in order to construct the
monopole gauge field as a coupling on the D-brane world-sheet. This
technique has been applied first in [9] for ADHM construction 
of instantons. In the present case the analysis will be somewhat
different due to the particularities of the model.

Perform a T-duality transformation of the previous configuration along
the $4567$ directions. 
\footnote {I thank M.R. Douglas for the idea of this analysis.}
The result is a system of two parallel type
IIb 7-branes in the $1,\ldots,7$ plane and $k$ parallel 5-branes in
the $4,\ldots,7,9$ plane stretching between them. The 5-branes intersect
the 7-brane along four dimensional subspaces with three transverse
coordinates $X^1,\,X^2,\,X^3$ within the 7-brane. The probe is a 
1-brane along the $9$-th axis also ending
\footnote{As pointed out by M.R. Douglas, the 1-brane cannot end on
the 7-brane since the $R-R$ 2-form does not couple to the
7-brane gauge field, so flux conservation would be violated. However,
the intersection point can be regarded for our purposes as an ending
point for the 1-brane fields. It is in this sense that we will be
using this terminology throughout the paper.}
on the 7-branes. Note that
this is exactly the configuration considered in [11] where it is
argued that the 5-brane is the monopole of the 7-brane world-volume.
The difference is that in this case the branes wrap the $T^4$ of
T-duality and not $K_3$, thus the 5-brane will be identified with a
monopole of N=4 rather than $N=2$ $SYM$ theory.
 
Quantization of different open string sectors yields:

$(i)$ Open $1-5$ strings have DD boundary conditions in $1238$ directions, DN
boundary conditions in $4567$ directions and NN boundary conditions in
$09$ directions. Quantization of the $NS$ sector yields four bosonic
states forming an $SO(4)\cong SU(2)\times SU(2)$ multiplet. GSO
projection selects $SO(4)$ chirality leaving a complex doublet.
Quantization of the $R$ sector yields chiral Weyl fermions $\xi_-$ and
$\xi_+$ with an internal symmetry $SO(4)$ group. GSO projection
relates world-sheet and internal chirality [9], leaving two $SU(2)$
doublets ${\xi_-}^A$, ${\xi_+}^Y$. The fields are charged under both the
1-brane $U(1)$ gauge field and the 5-brane $U(k)$ gauge field,
transforming as $(k,\,-1)\oplus ({\bar k},\,1)$. Moreover one expects
a mass-term for these fields proportional to $|X^i-Y^i|$, where $X^i$,
$Y^i$, $i=1,2,3$ describe
transverse oscillations of the five and one branes respectively.

$(ii)$ Open $1-7$ have DN boundary conditions along $1,\ldots,7,9$
directions, NN boundary conditions along $0,8$ directions and DD 
boundary conditions along the $9$-th direction. There is one
chiral fermion mode arising from the $R$ sector for each connected
component of the intersection between the 1-brane and the two branes. 
Thus we are left with two chiral fermions which constitute quantum
mechanical degrees of freedom located at different 1-brane endpoints.
These fields are expected two play an important role in the discussion
conditions for the bulk $1-5$ theory.

To derive the latter, we start as in [9] with five and nine branes and
then do dimensional reduction. More precisely, we start with $k$ 9-branes
and a 5-brane in the $012389$ plane and then take dimensional reduction
to the $09$ plane. The effective theory of the 5-brane is $D=6$
$(0,\,2)$ super Yang-Mills coupled to charged multiplets consisting of
a Weyl fermion $\xi$ and a doublet of complex scalars. The fermion
kinetic term is 
\be
{\cal L}_{kin}=\int d^6x\  {i{\bar\xi}{\Gamma}\cdot(\partial+A+iB)\xi}
\ee
where $A$ is the $U(k)$ gauge field induced from the 9-branes and
$B$ is the Abelian gauge field of the 5-brane.
The dimensional reduction is performed in [5]. The six dimensional
gauge fields yield two dimensional gauge fields and four scalars
representing oscillations of the 1-brane within the five brane. Since
the 1-brane is constrained to end on the 7-branes one of these
oscillations is frozen. Recall that $X^i$, $Y^i$, $i=1,2,3$ denote the
oscillations of the five and one brane respectively. 
The six dimensional Weyl fermion yields 
precisely the chiral complex fermions derived earlier from string 
quantization.
The relevant part of the two dimensional lagrangian for fermions is:
$$
{\cal L}^{(2)}=\int d^2x\
\{i{{\bar\xi}_-}D_+{\xi_-}+i{{\bar\xi}_+}D_-{\xi_+}\}
$$
\be
-\int d^2x\ \{i{\bar\xi}_+{(X^i+iY^i)\sigma_i}\xi_- +i{\bar\xi}_-{(X^i+iY^i)\sigma_i}\xi_+\}
\ee
Here $\sigma^i$ denote the standard $SU(2)$ generators acting on the
internal symmetry indices of the fermions. Note that because of GSO
projection the $SU(2)$ index of left handed fermions is different from
that of right handed fermions, thus the $\sigma^i$ carry 
mixed indices, $\sigma^i_{AY}$, and should not be thought of as
generators of any
of the $SU(2)$ groups in question. In the following we will simply
forget this subtlety and treat both $A$ and $Y$ on equal footing as
only one index with values $1,\,2$. Then we can define:
\be
\psi={1\over 2}(\xi_-+i\xi_+),\qquad \chi={1\over 2}(\xi_--i\xi_+)
\ee
and rewrite the lagrangian in the form:
$$
{\cal L}^{(2)}=\int d^2x\ \{i\bar\chi D_0 \chi + i\bar\psi D_0 \psi + 
i\bar\chi D_1 \psi + i\bar\psi D_1 \chi \}
$$
\be
- \int d^2x\ \{\bar\chi {(X^i+iY^i)\sigma_i}\psi -
\bar\psi{(X^i+iY^i)\sigma_i}\chi\}
\ee
Treating the terms without time derivatives as generalized mass terms
 we derive a quantum mechanical effective
action for fermions in the spirit of [37]. Since the gauge fields will
play no role in the following we can gauge them to zero.
The mass-less modes for the fermions are determined by the equations:
\be
(\partial_1 + (iX^i-Y^i)\sigma_i)\psi=0
\ee
\be
(\partial_1 - (iX^i-Y^i)\sigma_i)\chi=0
\ee
The first equation is identical to the equation (15) appearing in the 
construction of the monopole gauge field while the second equation is
it's dual. We will assume that the first equation has $p$ normalizable
solutions $v^\alpha$ analytic and finite near the boundary while the second
equation has none. This is the case if the fields $X^i$ satisfy either
Nahm or standard boundary conditions, but $p$ is different in each
case. 
Consider then the following low energy ansatz for the fermionic fields:
\be
\psi(t,s)=\psi^\alpha(t)v_\alpha(s),\qquad \chi(t,s)=0
\ee
where $\psi^\alpha(t)$ are slowly varying functions of time. Take
similarly $Y^i\equiv Y^i(t)$ to be slowly varying functions of time,
with no spatial dependence. Then the above lagrangian reduces to:
\be
{\cal L}^{(2)}=\int dt\ 
  {{\bar\psi}^\alpha}(\delta_{\alpha\beta}\partial_0+\partial_0 Y^i\left\{\int ds\
{{v_\alpha}^\dagger}\frac{\partial {v_\beta}}{\partial Y^i}\right\})\psi^\beta
\ee
We see that the probe moves in an $SU(p)$ space-time external
background gauge field ${\cal A}_j$ given by the Nahm construction
(17) where $p$ is the number of $1-5$ independent zero modes.
At this stage both Nahm and standard boundary conditions seem to lead
to a consistent theory, the probe moving in a  well defined monopole
gauge field. However one expects only one consistent theory and this
is where the $1-7$ boundary fields come into play.

\subsection{\it Boundary Conditions}

A proper analysis of the boundary $1-7$ fields and their couplings
with the bulk $1-5$ theory seems a rather difficult task, well
beyond the limits of this paper. We will not provide 
a complete solution here but we propose a  simple physical mechanism
powerful enough to select the boundary conditions.

The key observation is that the $1-5$ zero modes determined by 
Nahm equations are present even in the limit of large separation 
between the 1-brane and the 5-brane. The only mechanism that can make
this fact possible is to think of these modes as corresponding to 
$1-5$ strings gliding and touching the 7-branes (see fig.1). 

\vspace{5mm}
\centerline{\hbox{\psfig{figure=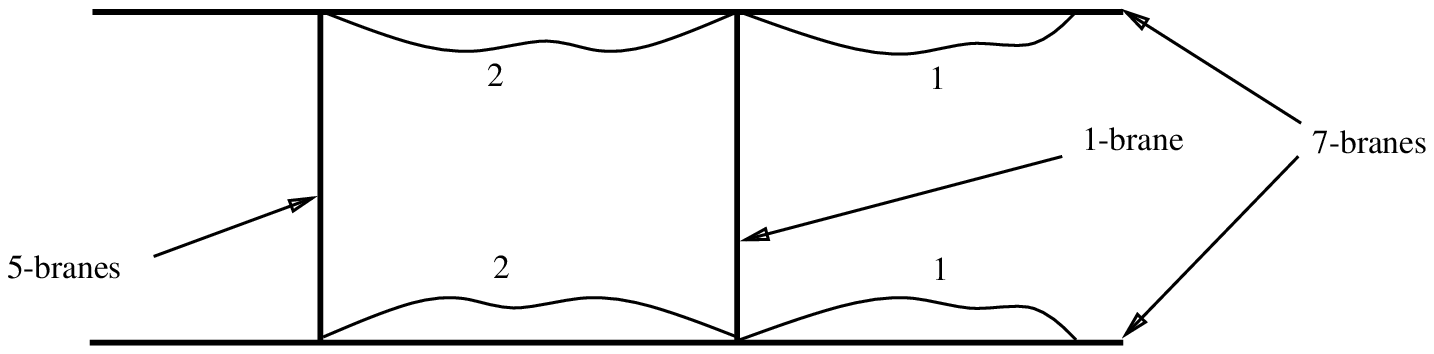}}}
\centerline{ \small {Figure1: The D-brane configuration for the probe
analysis}}
\centerline{ \small {$1$: $1-7$ tensionless strings}}
\centerline{ \small {$2$: $1-5$ strings that become
tensionless by touching the 7-branes}}
\vspace{5mm}

In this limit, the $1-5$ strings can be regarded as $1-7$ strings as well, 
thus there should be a $1-1$ correspondence between $1-5$ and $1-7$
zero modes. According to our analysis, there are only two 
$1-7$ zero modes located at the two endpoints of the 1-brane, thus
there should be only two zero modes in the $1-5$ theory for any value
of the 5-brane multiplicity $k$ i.e. $p=2$. 
This is all we need since that happens precisely when the ground state
fields $X^i$ satisfy Nahm boundary conditions! It is easy to check
that standard boundary conditions give $p=2k$ independent solution
leading thus to an inconsistency for any $k>1$. The exception $k=1$
is not a flaw in the argument as in that case they coincide [24].

A different aspect of this problem is revealed by the particular form
of the would be monopole gauge field ${\cal A}_j$. Trivial boundary
condition lead to an $SU(2k)$ gauge field, in fact a linear
superposition of embedded BPS monopoles. The probe analysis shows
that this is not physical since one would need the Chan-Paton factors 
of $2k$ 7-branes to construct it. Actually we have re-derived in the
D-brane framework an well known result in the monopole theory.
A linear superposition of $k>1$ BPS monopoles is not an exact 
$SU(2)$ monopole solution of charge $k$ [20], but it can be an exact
solution when embedded along different roots in a gauge group of
higher rank [20], [33], [34].

This shows that the natural ground states for the 1-brane
configurations considered so far are precisely those defined by Nahm
equations, which appear already in covariant form in this context. Defining 
complex Nahm data, it is straightforward to rewrite them
as moment map equations for an $U(k)$ hyperkahler quotient, as
expected in theories with extended super-symmetry. 

The irreducibility condition in the definition of Nahm data corresponds to
the fact that the $k$ 1-branes form a true bound state. Relaxing this
condition, we obtain a degenerate configuration corresponding to
infinitely separated sub-bound states as in the end of (2.2). 
This represents a point on the 
{\it boundary} of the monopole moduli space [1]. The usual flat
directions of the potential, while inconsistent when the branes are at 
finite distance, are recovered in the limit of infinitely separated
1-branes. Equivalently these are points on the boundary of the moduli
space parameterizing infinitely separated BPS monopoles.

Finally, there is an intriguing aspect of this picture, namely the 
Nahm ground states do not admit a clear coordinate interpretation
as in [35]. This is related to the fact that there is no general way
to define the centers of the hypothetical elementary magnetic charges
composing a multi-monopole solution [3],[4]. This can be done only for
well separated monopoles [1],[3],[4], that is in the asymptotic
region of the moduli space when one can show that the Higgs field has
exactly $k$ essential zeros (counted with multiplicity). The positions
of the zeros can be regarded as centers of magnetic charge. However,
this description breaks at a general point on the moduli space [17].
\footnote{ I thank R. Bielawski for clarifying these points to me. }

\section{D-Branes and $SU(n)$ Monopoles}

We generalize the results of the previous section to $SU(n)$
monopoles. The Nahm construction for $SU(2)$ monopoles can be
generalized to arbitrary gauge group as follows.

\subsection{\it SU(n) Monopoles and Nahm Equations}

The ADHM construction of arbitrary monopole moduli spaces has been
first discussed by Nahm, [23] and further developed in [18]. The brief
presentation in this subsection follows closely [18].
The asymptotic conditions for the Higgs field generalize to:
\be
\Phi(r)=i\mbox{diag}(\mu_1,\ldots\mu_n)-{i\over{2r}}\mbox{diag}
(k_1,{\ldots}k_n)+O({1\over{r^2}}) 
\ee
where $\mu_1\ldots\mu_n$ and $k_1{\ldots}k_n$ satisfy:
\be
\sum_{a=1}^n{\mu_a}=\sum_{a=1}^n{k_a}=0{,}\qquad
{\mu_1<\mu_2<\ldots <\mu_n}
\ee
The gauge group $SU(n)$ is asymptotically broken to a maximal torus 
$T=U(1)\times\ldots\times{U(1)}$ and the solutions are topologically
classified by homotopy classes ${[{\Phi}_{\infty}]}\in{\pi_2}(G/T)$.
According to [16], [22] these can be represented by 
$r$-tuples of integers $(m_1\ldots{m_r})$, where $r$ is the rank of the
group $G$, ${r=n-1}$ for $SU(n)$. The magnetic charges
$(m_1\ldots{m_{n-1}})$ are given in terms of the asymptotic data by:
\be
m_1=k_1{,}\qquad  m_2=k_1+k_2{,}\qquad  m_{n-1}=k_1+\ldots{k_{n-1}}
\ee
The Nahm data are defined as analytic $u(m_a)$ valued functions
$X^i_a$ defined on each interval $(\mu_a,\ \mu_a+1)$ such that they
solve Nahm equations:
\be
 {d{{X^i}_a}\over{ds}}+{1\over{2}}{\sum_{j{,}k=1}^3}\epsilon_{ijk}{[{X^j}_a,
{X^k}_a]}=0  
\ee
subject to the boundary and matching conditions:

$(i)$ Let $t=s-\mu_a$. Then at a point $\mu_a$ with $m_{a-1}<m_a$,
${X^i}_{a-1}$ is analytic and has finite nonzero limit ${C^i}_a$ as
$t\mapsto{0^-}$ and ${X^i}_a$ has a block form expansion near $t=0$
\be
  \begin{array}({cc})
       {C^i}_a+O(t) & O(t^{\gamma}) \\
       O(t^{\gamma}) & {{T^i}_a\over{t}}+O(1)
  \end{array}
\ee
where $\gamma={{m_a-m_{a-1}-1}\over{2}}$ and ${X^i}_a$ define an
irreducible representation of $SU(2)$.

$(ii)$ If $m_{a-1}>m_a$ the boundary conditions are the same with the
roles of $m_a$ and $m_{a-1}$ inversed.

$(iii)$ If $m_{a-1}=m_a$, ${X^i}_a$ and ${X^i}_{a-1}$ are both analytic
near $t=0$ with 
finite limits $C^{i+}_a$ and $C^{i-}_a$ required to satisfy
a certain regularity condition which will not be made explicit here
since it will play no role in the following. A stronger version of
that condition is simple continuity $C^{i+}_a=C^{i-}_a$ and it restricts us
in this case to embedded $U(n-1)$-monopoles [18]. 

Note that index $a$ runs from $0$ to $n$ with the conventions
$m_0=m_n=0$.
Then there is a $1-1$ correspondence between $U(m_a)$ conjugacy classes
of Nahm data and $SU(n)$ monopoles with magnetic charges
$(m_1,\ldots,m_{n-1})$. Repeating the steps of the $SU(2)$
construction on each interval one can define complex Nahm data and
represent the moduli space as a hyperkahler quotient. 
Taking into account the boundary conditions for
${X^i}$ the complex gauge transformations are properly defined [18] as
$(n-1)$-tuples $g=(g_1{,}\ldots{g_{n-1}})$ of smooth maps
$g:(\mu_{a-1}{,}\ {\mu_a})$ with finite analytic limits at the
boundary such that:

$i)$ If $m_a>{m_{a-1}}$, $g_a$ preserves the block form (56), the
derivatives of the off diagonal blocks are of order $O(\gamma)$ and
the limit of the ${m_a}\times{m_a}$ diagonal block equals the limit of
$g_{a-1}$.

$ii)$ If $m_a<{m_{a-1}}$ the same boundary conditions are satisfied
with roles of $m_a$ and $m_{a-1}$ inversed.

$iii)$ If $m_a={m_{a-1}}$ the limits of $g_a$ and $g_{a-1}$ coincide.

\subsection{\it D-Brane Configurations and Moduli Spaces}

The construction of an $SU(n)$ monopole with asymptotic conditions
(52), (53) requires a configuration of $n$ parallel 3-branes located
at points of coordinates:
\be
X^9_a=\mu_a,\qquad a=1,\ldots,n
\ee
along the $x^9$ axis.
This configuration corresponds to a point of
maximal symmetry breaking of the effective $U(n)$ gauge theory on the
3-brane world-volume. The gauge group may be factored as $U(n)\cong
SU(n)\times U(1)$ where the Abelian factor describes the center of
mass dynamics of the 3-branes and the $SU(n)$ factor describes the
relative dynamics of the branes. The subsequent construction will thus
yield $SU(n)$ monopoles rather than $U(n)$ monopoles. 

\vspace{5mm}
\centerline{\hbox{\psfig{figure=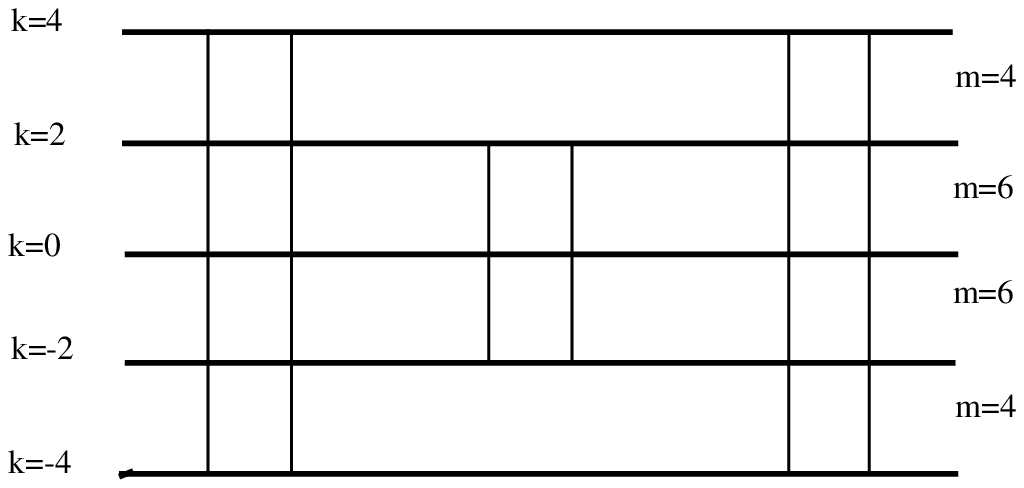}}}
\centerline{\small Figure2: The D-brane configuration for a
spherically symmetric $SU(5)$ monopole}
\vspace{5mm}

The key is
to construct a configuration of 1-branes with endpoints on
the $n$ parallel 3-branes so that the net magnetic charge induced on
the $a$-th brane be equal to $k_a$ and that in the interval $(\mu_a{,}\
\mu_{a+1})$ there be exactly $m_a$ 1-branes stretching along the
$x^9$ axis. The best way to illustrate how this works is to present
some particular cases which will make the general rules clear. The
details for spherically symmetric $SU(5)$ and $SU(6)$
monopoles are presented in figures one and two respectively.

\vspace{5mm}
\centerline{\hbox{\psfig{figure=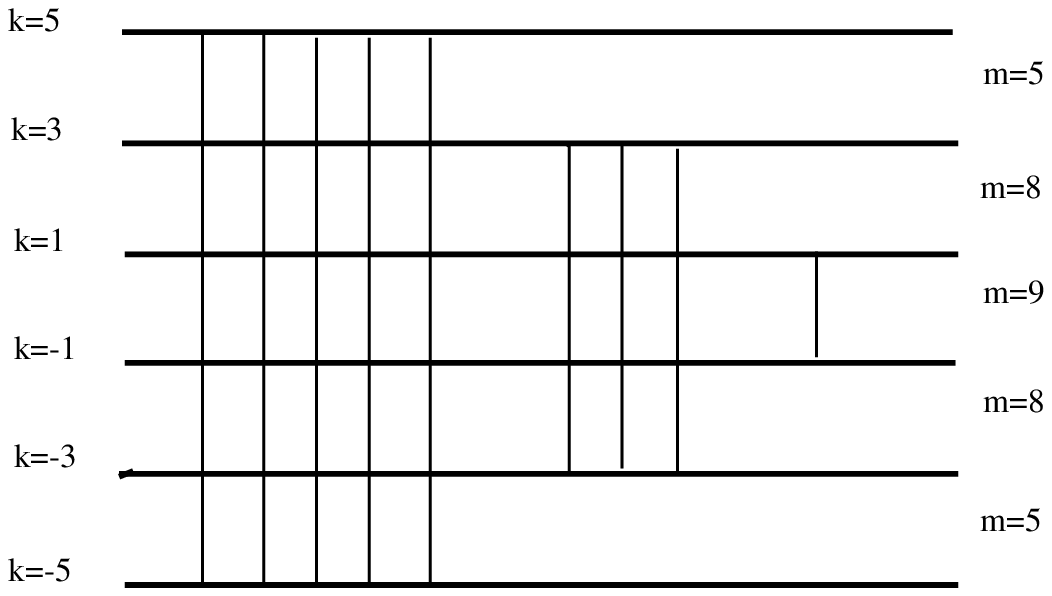}}}
\centerline{\small Figure3: The D-brane configuration for a
spherically symmetric $SU(6)$ monopole}
\vspace{5mm}

Standard Gauss law arguments show that a 1-brane ending on a pair
of 3-branes induces a monopole on one of the branes and an
anti-monopole on the other. Moreover a 1-brane threading through the
core of a 3-brane does not induce any charge in the 3-brane
world-volume. Consequently, the oscillations of the 1-brane are not
constrained by charge conservation, the world-sheet fields being continuous.

As in the case of $SU(2)$ monopoles, boundary and matching conditions
play a crucial role in the construction. In fact, a part of these are
already implicit in the previous paragraph. For a systematic approach,
let the $u(m_a)$ valued world-sheet fields $X^i_a$ describe the
transverse oscillations of the 1-branes stretching between the $a$-th
and the $(a+1)$-th 3 brane. Charge conservation yields 
$m_a=m_{a-1}+k_a$, where $k_a$ is the net magnetic charge on the
$a$-th 3-brane,  equal to the net number of 1-branes that end on that
brane. The Chan-Paton degrees
of freedom of the lower configuration can be split as:
\be
C^{m_a}=C^{m_{a-1}}\oplus C^{k_a}
\ee
and the matrix valued fields $X^\mu,\ X^i$ can be set in block form:
\be
    {X^\mu}_a={\begin{array}({cc})
            {X^\mu}_{a,11} & {X^\mu}_{a,12} \\
            {X^\mu}_{a,21} & {X^\mu}_{a,22}
            \end{array}}\qquad
    {X^i}_a={\begin{array}({cc})
            {X^i}_{a,11} & {X^i}_{a,12} \\
            {X^i}_{a,21} & {X^i}_{a,22}
         \end{array}}
\ee
The fermionic fields can be split similarly:
\be
    {\chi^A}_a={\begin{array}({cc})
            {\chi^A}_{a,11} & {\chi^A}_{a,12} \\
            {\chi^A}_{a,21} & {\chi^A}_{a,22}
            \end{array}}
\ee
The boundary conditions for the fields $X^\mu$, $\mu=4,\ldots,8$ are:

$(i)$ ${X^\mu}_{a-1}$, ${X^\mu}_{a,11}$ must be analytic in a neighborhood of
$t=0$ with finite limits as $t\mapsto 0$ and these limits must
coincide.

$(ii)$ ${X^\mu}_{a,12}$, ${X^\mu}_{a,21}$, ${X^\mu}_{a,22}$ must be ``bump
fields'' that is compactly supported away from $t=0$.

Note that the condition $(i)$ above holds for the fields $X^i$ as
well. The matching conditions for fermions are identical. The surface
terms arising from the super-symmetry variation of the upper and lower 
lagrangian cancel each other leading again to a consistent theory.

Imposing the reality condition (24) on each interval we find a family
of super-symmetric ground states which are solutions to covariant Nahm 
equations:
\be
D_1{X^i}_a+{1\over 2}\epsilon^{ijk}[{X^j}_a,\ {X^k}_a]=0
\ee
We fix again the gauge $A_1=0$ reducing to standard Nahm equations. In
order to analyze the local behavior of the solutions, rewrite these
equations in terms of matrix blocks:
$$
\frac{d{X^i}_{11}}{ds}+{1\over 2}\epsilon^{ijk}([{X^j}_{11},\ {X^k}_{11}]+{X^j}_{12}{X^k}_{21}-{X^k}_{12}{X^j}_{21})=0
$$
$$
\frac{d{X^i}_{12}}{ds}+{1\over
2}\epsilon^{ijk}({X^j}_{11}{X^k}_{12}-{X^k}_{11}{X^j}_{12}+{X^j}_{12}{X^k}_{22}-{X^k}_{12}{X^j}_{22})=0
$$
$$
\frac{d{X^i}_{21}}{ds}+{1\over2}\epsilon^{ijk}({X^j}_{21}{X^k}_{11}-{X^k}_{21}{X^j}_{11}+{X^j}_{22}{X^k}_{21}-{X^k}_{22}{X^j}_{21})=0
$$
\be
\frac{d{X^i}_{22}}{ds}+{1\over 2}\epsilon^{ijk}([{X^j}_{22},\
{X^k}_{22}]+{X^j}_{21}{X^k}_{12}-{X^k}_{21}{X^j}_{12})=0
\ee
where the interval index $a$ has been suppressed for simplicity. The
matching conditions for fermions imply by consistency with the
unbroken symmetries that:

$(i)$ ${X^i}_{11}$, ${X^i}_{a-1}$ must be analytic in a neighborhood of
$t=0$, with finite and equal limits as $t\mapsto 0^+$ and $t\mapsto
0^-$ respectively.

$(ii)$ ${X^i}_{12}$, ${X^i}_{21}$ must be analytic
neighborhood of $t=0$, vanishing to some order $\gamma$ as $t\mapsto
0$.

$iii)$ ${X^i}_{22}$ must be identically equal near $t=0$ to a general
local solution to Nahm equations:
\be
{X^i}_{22}={{T^i}\over t}+O(1)
\ee
where $T^i$ define an irreducible representation of $SU(2)$.

The vanishing order $\gamma$ can be determined from representation
theoretic consistency considerations similar to those of [13], pg. 625.
The result is 
\be 
\gamma= {{k_a-1}\over 2}
\ee
being identical to that in (56).
Finally, if $m_{a-1}=m_a$ the above conditions reduce simply to:
\be
X^i_{a-1}=X^i_a,\qquad t=0
\ee
which is a stronger condition than that of [13] for monopole moduli
spaces and it corresponds to an embedded $U(n-1)$ monopole. This
appears quite naturally in the D-brane picture since $k_a=0$ implies
that there are no magnetic charges induced on the $a$-th 3-brane, thus
it can be removed from the configuration with no effect on the charge 
distribution. The presence of the poles in the boundary conditions of
the transverse fields has been analyzed in detail in the previous section. 

In conclusion, the D-brane-monopoles correspondence can be generalized
to $SU(n)$ gauge groups with arbitrary $n$. We remark that the moduli
space of super-symmetric ground states are naturally described by
covariant Nahm equations. The formulation in terms of hyperkahler
quotients is again straightforward. One has to rearrange the Nahm data
in complex form and to define appropriate gauge transformations. Since
we have shown that the boundary conditions for the fields are
precisely Nahm conditions, it is clear that the gauge transformations
are identically to those presented in (3.1) for monopoles.
A distinct feature of the above generalization is that the D-brane 
configurations admit a natural interpretation in terms of BPS
monopoles embedded along different roots. This will be pursued next.

\subsection{\it Embedded BPS Monopoles in D-Brane Picture}

Embedding of elementary monopole solutions in arbitrary gauge groups 
has been thoroughly studied in [20], [33], [34]. Following this line,
one can choose an orthonormal basis of the
$su(n)$ Cartan sub-algebra such that the asymptotic value of the Higgs
field takes the form:
\be
\Phi={\bf{\mu}}\cdot {\bf{H}}-{1\over r}{\bf{k}}\cdot{\bf{H}}+O({1\over r^2})
\ee
In the case of maximal symmetry breaking, the simple roots
$\beta^a$, $a=1,\ldots,n-1$ can be uniquely chosen so that:
\be
\bf{\mu}\cdot\bf{\beta^a}>0
\ee
The simple roots can be described conveniently as vectors
in an $n$-dimensional space with basis $\{{\bf{e}}_i\}$ lying in the
hyper-plane perpendicular to $\sum_{i=1}^n{{\bf{e}}_i}$:
\be
{\bf{\beta}}^a={\bf{e}}_a-{\bf{e}}_{a+1}
\ee
An arbitrary root $\bf{\alpha}$ defines
an $SU(2)$ subgroup with generators:
$$
t_1={(2{\bf{\alpha}}^2)}^{-1/2}(E_\alpha+E_{-\alpha})
$$
$$
t_2=-i{(2{\bf{\alpha}}^2)}^{-1/2}(E_\alpha-E_{-\alpha})
$$
\be
t_3={\bf{\alpha}^*}\cdot\bf{H}
\ee
where 
\be
\bf{\alpha}^*={{\bf{\alpha}}\over{\bf{\alpha}^2}}
\ee
is the dual of ${\bf\alpha}$.
Then one can embed a fundamental $SU(2)$ monopole in $SU(n)$ along
any root ${\bf\alpha}$ in a natural way [20], [33], [34]. If
${\bf{\alpha}}\equiv{\bf{\beta}}^a$ is a simple root the resulting
solution has magnetic charges:
\be
m_{ab}=\delta_{ab}
\ee
These are called fundamental $SU(n)$ monopoles. If the root
$\bf\alpha$ is not simple the solution can be regarded as a
superposition at the same point in space of fundamental solutions 
oriented along different simple
roots . The magnetic charges of such a
solution are given by:
\be
{\bf{\alpha}}^*=\sum_{a=1}^k{m_a}{\bf{\beta}^*_a}
\ee
Note that in the case of non-simple roots the superposition described
above is still an exact solution. This is not true for $SU(2)$ gauge
group, leading to the contradiction discussed in section 2.4.
Approximate solutions with higher charges can be obtained
by embedding many well separated fundamental monopoles along the same
simple root, similarly to the construction of the asymptotic region of
the moduli space of $SU(2)$ monopoles.

Returning to D-brane configurations, note that there is a $1-1$
correspondence between the set of vectors $\{{\bf{e}}_i\}$ and the set
of $n$ parallel 3-branes and that each pair of consecutive 3-branes
$(a,\ a+1)$ determines uniquely a simple root.
Then, taking into
account the results of the previous section, a 1-brane stretching
between these two 3-branes can be identified with the fundamental
$SU(2)$ monopole embedded along the corresponding simple root. Many
widely separated 1-branes in the same position can be identified with
an asymptotic superposition of fundamental monopoles. A 1-brane
stretching between two non-consecutive 3-branes $(a,\ b)$, $a<b-1$ can be
identified with an $SU(2)$ monopole embedded along the root:
\be
{\bf{\alpha}}=({\bf{e}}_a-{\bf{e}}_{a+1})+\ldots+({\bf{e}}_{b-1}-{\bf{e}}_{b})
\ee
that is to a superposition of fundamental monopoles corresponding to
${\bf{\beta}}_a$, $\ldots$, ${\bf{\beta}}_{b-1}$ at the same point in
space. This agrees with the representation of the $(a,\ b)$ 1-brane as
a collection of 1-branes $(a,\ a-1),\ldots,(b-1,\ b)$ with endpoints
identified. Note also that since 
\be
{\bf{\beta}}^*_a={1\over 2}{\bf{\beta}}_a,\qquad{\bf{\alpha}}^*={1\over
2}{\bf{\alpha}}
\ee
the magnetic charges of this solution are:
\be
m_1=0,\ldots\ ,m_{a-1}=0,\ m_a=1,\ldots\ ,m_{b-1}=1,\ m_b=0\ldots\ ,m_{n-1}=0 
\ee 
in perfect agreement with those of the D-brane configuration. Many
widely separated $(a,\ b)$ 1-branes would yield an asymptotic
multi-monopole configuration as above. This picture suggests that an
arbitrary bound state of 1-branes can be interpreted similarly as a
generic $SU(2)$ monopole embedded in $SU(n)$ yielding an exact $SU(n)$
solution. It is not clear if this holds true.

\section{Conclusion}

We have shown that the identification of the 3-brane world-volume
monopoles with the D-strings of the type IIb theory can be made
explicit in Nahm formalism. The detailed study of this correspondence 
has revealed several new aspects of D-brane physics.
Perhaps the most intriguing of all is the coordinate interpretation
of the Nahm super-symmetric ground states. Consistency arguments have
shown that the transverse fields $X^i$ describing the positions of the
1-brane endpoints within the 3-brane develop poles at the boundary! 
Moreover while they are smooth on the interior, they do not commute,
thus a direct coordinate interpretation as in [35] is missing.
This fact is related to the fact that monopoles are massive objects
which generally do not have a well defined location in space. Since
the 1-brane endpoints may be considered point-like particles, the poles
appear as an attempt at a reconciliation between these two aspects.
This provides further evidence that D-branes are more than simple 
geometrical objects, their behavior in certain circumstances 
contradicting standard geometrical interpretation. The consequences,
as well as the extent of this phenomenon remain rather mysterious.

Another interesting aspect, not emphasized in the text, is the 
hidden correspondence between instanton/monopole reciprocity, [6],
and D-branes. This fact has been first noted in [10] where it is
argued that instanton reciprocity can be regarded as a T-duality 
transformation which 
interchanges outer and inner quivers. Since the 3-brane and the
1-brane can also be interchanged by a T-duality transformation, the
above results show that this correspondence extends to monopole
moduli space. Note also that in the case of instantons, reciprocity 
transforms the self-duality equations in pure algebraic equations
which describe the moduli space of a Dirichlet $p$-brane embedded
in a $p+4$ brane. In the present case reciprocity transforms the 
self-duality equations in a system of ordinary differential equation,
corresponding to the fact that the 1-brane is transverse to the
3-brane. Finally it is interesting to note that performing a T-duality
transformation along the 1-brane we end up with a system of type IIa
$0$-branes embedded in a 4-brane whose moduli space should 
describe instantons. It appears that the T-duality interchanges
monopoles and instantons! Although not very clear at the present stage
this line of development might lead to new insights in the interplay
between D-branes and moduli spaces of solitonic objects.

\newpage
{\bf{References}}
\begin{list}{}
\item {[1] M.F. Atiyah and N.J. Hitchin {\it The Geometry and Dynamics
of Magnetic Monopoles}, Princeton Univ. Press, Princeton (1988)}
\item {[2] M.C. Bowan, E. Corrigan, P. Goddard, A. Puaca and A. Soper
{\it Construction of Spherically Symmetric Monopoles Using the
Atiyah-Drinfeld-Hitchin-Nahm Formalism}, Phys. Rev. D28, 3100 (1983)} 
\item {[3] R. Bielawski, {\it Monopoles, Particles and Rational
Functions}, McMaster preprint (1996), to appear in Ann. Glob. Anal. Geom.}
\item {[4] R. Bielawski, {\it Asymptotic Behaviour of SU(2) Monopole
Metrics}, J.Reine.Angew.Math. 468, 139 (1995)}
\item {[5] L. Brink, J. Schwarz and J. Scherk, {\it Supersymmetric
Yang-Mills Theories}, Nucl. Phys. B121, 77 (1977)}
\item {[6] E. Corrigan and P. Goddard, {\it Construction of Instanton
and Monopole Solutions and Reciprocity}, Ann. Phys. 154, 253 (1984)}
\item {[7] J. Dai, R.G. Leigh, J. Polchinski, {\it New Connections
Between String Theories}, Mod. Phys. Lett. A4, 2073 (1989)}
\item {[8] S.K. Donaldson, {\it Nahm Equations and the Classification
of Monopoles}, Commun. Math. Phys. 96, 387 (1985)}
\item {[9] M.R. Douglas, {\it Gauge Fields and D-Branes},
hep-th/9604198}
\item {[10] M.R. Douglas and G. Moore, {\it D-Branes, Quivers and ALE
Instantons}, hep-th/9603167}
\item {[11] M.R. Douglas and M. Li, {\it D-brane Realization of N=2
Super Yang-Mills Theory in Four Dimensions}, hep-th/9604041}
\item {[12] M.B. Green and M. Gutperle, {\it Comments on D-Branes}, hep-th/9604091}
\item {[13] J.A. Harvey, {\it Magnetic Monopoles, Duality and
Supersymmetry}, hep-th/9603086}
\item {[14] N.J. Hitchin, {\it Hyperkahler Quotients}, Asterisque 206,
137 (1992)}
\item {[15] N.J. Hitchin, {\it On the Construction of Monopoles},
Commun. Math. Phys. 89,145 (1983)}
\item {[16] P.A. Horvathy and J.H. Rawnsley, {\it Topological Charges
in Monopoles Theories}, Commun. Math. Phys. 96, 497 (1984)}
\item {[17] C.J. Houghton and P.M. Sutcliffe, {\it Monopole Scattering
With a Twist}, Nucl.Phys.B 464, 59 (1996)}
\item {[18] J. Hurtubise, {\it The Classification of Monopoles for the
Classical Groups}, Commun. Math. Phys. 120, 613 (1989)}
\item {[19] J. Hurtubise, {\it Monopoles and Rational Maps: A Note on
a Theorem of Donaldson}, Commun. Math. Phys. 100, 191 (1985)}
\item {[20] K. Lee, E.J. Weinberg and P. Yi, {\it The Moduli Space of
Many Monopoles for Arbitrary Gauge Groups}, hep-th/9602167}
\item {[21] R.G. Leigh {\it Dirac-Born-Infeld Action From Dirichlet
Sigma Model}, Mod. Phys. Lett. A4, 2767 (1989)}
\item {[22] M.K. Murray {\it Non-Abelian Magnetic Monopoles},
Commun. Math. Phys. 96, 539 (1984)}
\item {[23] W. Nahm, {\it The Construction of All Self-Dual
Multimonopoles by the ADHM Method}, ``Monopoles in Quantum Field
Theory'', Craigie et al. (eds), World Scientific, Singapore (1982) }
\item {[24] W. Nahm, {\it A Simple Formalism for The BPS Monopole},
Phys. Letters 90B, 413 (1980)}
\item {[25] H. Nakajima, {\it Monopoles and Nahm Equations}, Einstein
Metrics and Yang-Mills connections (Sanda,1990),193, Lecture Notes in
Pure and Applied Mathematics, 145, Dekker, New York (1993)}
\item {[26] J. Polchinski, S. Chauduri, C.V. Johnson, {\it Notes on
D-Branes}, hep-th/9602052}
\item {[27] J. Polchinski, {\it Dirichlet Branes and Ramond-Ramond
Charges}, Phys. Rev. Lett. 75, 4724 (1995)}
\item {[28] J. Polchinski and E. Witten, {\it Evidence for
Heterotic-Type I String Duality}, Nucl. Phys. B460, 525 (1996)}
\item {[29] C. Schimdhuber, {\it D-Brane Actions}, hep-th/9601003}
\item {[30] J.H. Schwartz, {\it An SL(2,\,Z) Multiplet of Type IIB
Superstrings}, hep-th/9508143}
\item {[31] A. Strominger, {\it Open P-Branes}, hep-th/9512059}
\item {[32] A.A. Tseytlin, {\it Selfduality of Born-Infeld Action and
Dirichlet 3-brane of Type IIB Superstring Theory}, hep-th/9602064}
\item {[33] E.J. Weinberg, {\it Fundamental Monopoles and
Multimonopole Solutions for Arbitrary Simple Gauge Groups},
Nucl. Phys. B167, 500 (1980)}
\item {[34] E.J. Weinberg, {\it Fundamental Monopoles in Theories with
Arbitrary Symmetry Breaking},
Nucl. Phys. B203, 500 (1982)}
\item {[35] E. Witten, {\it Bound States of Strings and p-Branes},
hep-th/9510135}
\item {[36] E. Witten, {\it Small Instantons in String Theory},
hep-th/9511030}
\item {[37] E. Witten, {\it Sigma Models and The ADHM Construction of 
Instantons}, hep-th/9410052}
\end{list}

\end{document}